# Phonon-enhanced superconductivity at the FeSe/SrTiO$_3$ interface

Q. Song[1,2,3,+], T. L. Yu[1,2,3,+], X. Lou[1,2,3], B. P. Xie[2], H. C. Xu[2], C. H. P. Wen[1,2,3], Q. Yao[1,2,3], S. Y. Zhang[4,5], X. T. Zhu[4,5], J. D. Guo[4,5,6], R. Peng[2,*], D. L. Feng[1,2,3,*]

[1] State Key Laboratory of Surface Physics and Department of Physics, Fudan University, Shanghai 200433, China

[2] Laboratory of Advanced Materials, Fudan University, Shanghai 200433, China

[3] Collaborative Innovation Center of Advanced Microstructures, Nanjing, 210093, China

[4] Beijing National Laboratory for Condensed Matter Physics and Institute of Physics, Chinese Academy of Sciences, Beijing 100190, China

[5] School of Physical Sciences, University of Chinese Academy of Sciences, Beijing 100049, China

[6] Collaborative Innovation Center of Quantum Matter, Beijing 100871, China

[+] These authors contributed equally.

**The dream of room temperature superconductors has inspired intense research effort to find routes for enhancing the superconducting transition temperature ($T_c$). Therefore, single-layer FeSe on a SrTiO$_3$ substrate, with its extraordinarily high $T_c$ amongst all interfacial superconductors and iron based superconductors, is particularly interesting, but the mechanism underlying its high $T_c$ has remained mysterious. Here we show through isotope effects that electrons in FeSe couple with the oxygen phonons in the substrate, and the superconductivity is enhanced linearly with the coupling strength atop the intrinsic**

**superconductivity of heavily-electron-doped FeSe. Our observations solve the enigma of FeSe/SrTiO$_3$, and experimentally establish the critical role and unique behavior of electron-phonon forward scattering in a correlated high-T$_c$ superconductor. The effective cooperation between interlayer electron-phonon interactions and correlations suggests a path forward in developing more high-T$_c$ interfacial superconductors, and may shed light on understanding the high T$_c$ of bulk high temperature superconductors with layered structures.**

General routes to enhance the superconducting transition temperature (T$_c$) of existing high T$_c$ materials are a long-sought goal in superconductivity research. Recent experiments on FeSe/SrTiO$_3$ (FeSe/STO) films suggest a greatly enhanced T$_c$ of 65 K or higher[1], compared to 8 K in bulk FeSe (ref. 2). The T$_c$ of 65 K was confirmed by converging data from angle-resolved photoemission spectroscopy (ARPES)[3-5], mutual inductance[6] and muon spin relaxation measurements[7] although a resistivity downturn has been observed as high as 109 K (ref. 8) . ARPES studies show that the T$_c$ in FeSe/STO/KTaO$_3$ and FeSe/BaTiO$_3$ systems can reach 70 K and 75 K, respectively[9,10]. In contrast, heavily electron-doped FeSe systems, whether the doping is introduced through liquid or solid gating[11-13], surface dosing[14,15], intercalation[16-18], or chemical substitution[19], superconduct at 46~48 K or below. The significant enhancement of T$_c$ in single-layer FeSe on oxide substrates is evidently related to the interface; however, the mechanism remains mysterious despite intensive study. Uncovering the mechanism of the remarkable interfacial superconductivity enhancement would set a path toward interfacially-enhanced superconducting systems and devices based on the known high T$_c$ superconductors.

Several scenarios have been proposed to address the crucial role of the interface. For example, interfacial tensile strain has been proposed to enhance the antiferromagnetic exchange interaction in FeSe, thus enhancing the superconductivity[20]. However, experiments exclude this scenario based on the negligible change of $T_c$ in films with varied strain[9,10,21,22]. Alternatively, interfacial electron phonon interactions (EPI) were suggested to enhance $T_c$, since echoes of the FeSe band structure (*i.e.* replica or shake-off side bands) have been observed shifted by an energy similar to those of optical phonons in STO[5,10,23], and pronounced Fuchs-Kliewer (FK) surface oxygen optical phonons have been observed on a single-layer FeSe/STO film by electron energy loss spectroscopy (EELS)[24]. However, no consensus has been reached on the role of the interfacial EPI or the magnitude of enhancement. Some theories suggest that the interfacial EPI must collaborate with electron correlations in heavily electron-doped FeSe to give the high $T_c$[5, 25-28], while others indicate that EPI alone can fully account for the 65~75 K $T_c$ (ref. 29). Still others argue that the EPI would be strongly screened and thus irrelevant to superconductivity[30], and some even suggest the side band observed in ARPES can be readily interpreted as a renormalized Fe $3d_{xy}$ band[31]. To settle the controversies on the role of EPI, and reveal the $T_c$-enhancement mechanism in FeSe/STO, direct experimental evidence is needed. Here we show O-isotope dependence on the shake-off side bands in FeSe/STO using ARPES and EELS, directly linking them with forward scattering between the electrons in FeSe and the oxygen optical phonons in the substrate. Moreover, the superconducting gap of FeSe/STO increases linearly with the side band intensity, while extrapolated to the limit of vanishing side band intensity it reaches the gap value of a heavily-electron-doped FeSe monolayer. This proves that the high $T_c$ in FeSe/STO is achieved jointly by the interfacial EPI and the intrinsic superconductivity of heavily-electron-doped FeSe.

**Phonon origin of the side bands**

To understand whether the correlation between the phonon energy and side-band offset from the main band is a coincidence or indeed due to the interfacial EPI, we start by studying how the side band is related to the STO phonons. Three types of electron-doped FeSe films were prepared, as sketched in Fig. 1a. The first is a 50 monolayer (ML) FeSe film grown on a ST$^{16}$O substrate. After sufficient potassium (K) dosing on its surface, the surface FeSe layer becomes heavily electron doped[14,15,32,33], with a $T_c$ of about 46 K[14]. The other two are single-layer FeSe films grown on 60 unit-cell films of ST$^{16}$O and ST$^{18}$O, respectively, on ST$^{16}$O substrates (see Methods and Supplementary Information for details), which are labelled as S16 and S18 hereafter. Their Fermi surfaces all consist of two ellipses in the two-iron Brillouin zone (see the dashed and solid ellipses in Fig. 1b)[9,34], one of which is much more pronounced here due to the experimental geometry[9]. Figures 1c and 1d show the photoemission intensities for these three types of films in their superconducting states along cut #1, and the corresponding second-derivatives with respect to energy, respectively. They exhibit essentially the same band structure with the more pronounced electron band noted as γ, which bends near the normal state Fermi momentum ($k_F$), signifying the opening of a superconducting gap. The main difference among them is the two replica bands, γ' and γ*, observed on the single layer FeSe/STO only, consistent with previous ARPES results on FeSe/STO (ref. 5). The replica band γ' has been attributed to the phonon shake-off mechanism due to the forward scattering of the electrons with an oxygen optical phonon[5], since it duplicates the main band *without any momentum shift*, and the separation between γ and γ' is close to the energy of an STO optical phonon[24]; however, this replica band has also been argued to be merely a renormalized $3d_{xy}$ band[31]. Its observation only in our monolayer FeSe films implies that the interface must play a crucial role in its appearance.

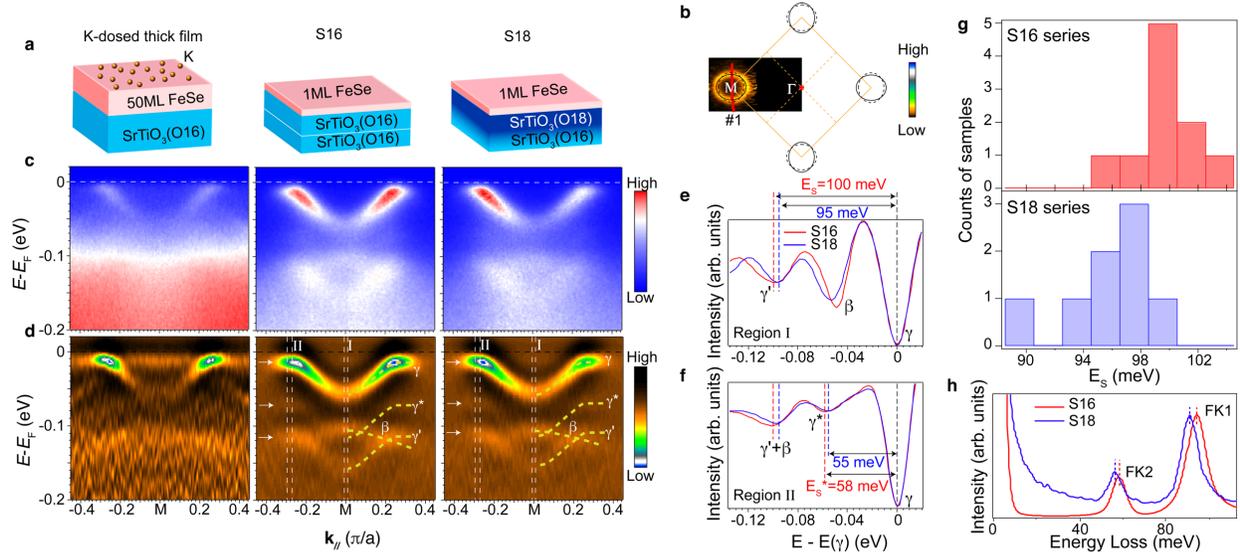

**Figure 1 Isotope dependence of the phonon features. a,** Sketch of three types of FeSe films grown on STO substrates: K-dosed thick FeSe film, single-layer FeSe/SrTi$^{16}$O$_3$ (referred to as S16), and single-layer FeSe/SrTi$^{18}$O$_3$ (referred to as S18). **b,** Photoemission intensity map at the Fermi energy and the sketch of Fermi surfaces of S16. **c,d,** Photoemission intensity (**c**) along cut #1 across M shown in panel b and the corresponding second derivative with respect to energy (**d**) to highlight the dispersions for the K-dosed thick FeSe film, S16, and S18. **e,f,** Integrated second derivative of the energy distribution curves (EDCs) for S16 and S18 with respect to the binding energy of band γ at momentum regions I (**e**) and II (**f**) indicated in panel (**d**). **g,** Histogram of the energy separations between band γ and γ' in various S16 and S18 samples. **h,** Energy loss spectra measured by high resolution EELS for S16 and S18. The photoemission data were measured at 12K, and the EELS data were measured at 35K.

To unveil the origin of these side bands, we quantitatively study how the band energy separations relate with the oxygen isotope effect on the phonon energies. Figures. 1e-f display the second derivative data integrated over the momentum regions indicated in Fig. 1d, for both S16 and S18.

The minima in the second derivative mark the band positions, which give energy separations between γ and γ' (noted as $E_S$) of 100 meV for S16, and 95 meV for S18, respectively (Fig.1e). To estimate the deviations from film to film, photoemission data were collected from eighteen samples, and statistical analysis (Fig. 1g) show that the $E_S$'s are 100±2meV and 95±3meV for S16 and S18, respectively, demonstrating the isotope dependence. $E_S$ in S18 films is about 5% less than that in the S16 films. This is close to the difference from unity of the square root of the ratio of the $^{18}O$ and $^{16}O$ masses (6%), suggesting that the side bands are related to oxygen phonons. A similar trend is observed in the isotope dependence of the γ∗ band, with energy separations between γ and γ∗, $E_S$∗, of 58 meV in S16 and 55 meV in S18 (Fig.1f). To resolve the interfacial phonons, we performed *ex-situ* EELS study on the samples after ARPES measurements. As shown in Fig.1h, the EELS spectra of both samples clearly show the features for FK1 and FK2 phonons of the STO substrates[24]. The FK1 energy, $\Omega_1$, is 94.2 meV in S16, and 91.0 meV in S18, while the FK2 energy, $\Omega_2$, is 58.3 meV in S16, and 56.4 meV in S18, respectively. $\Omega_1$ in S18 is 3.5% softer than its counterparts in S16, and $\Omega_2$ is 3.4% softer. The facts that $\Omega_1$ ($\Omega_2$) is close to $E_S$ ($E_S$∗), and that the same isotope dependences are observed in all energies constitute compelling evidence that the side bands γ' and γ∗ in single-layer FeSe/STO are due to the scattering from STO FK1 and FK2 phonons, respectively. The slight differences between $\Omega_1$ ($\Omega_2$) and $E_S$ ($E_S$∗) can be attributed to the renormalization of the bands due to EPI[5,29], as will be discussed later. The observed larger isotope response in $E_S$ ($E_S$∗) than that in $\Omega_1$ ($\Omega_2$) may be due to the fact that the EELS measurements were conducted *ex situ* after the ARPES measurements, and the additional heat treatment may cause oxygen diffusion between the ST$^{16}$O substrate and the ST$^{18}$O film (see Methods for more details of EELS measurements), making it partially substituted. Both FK1 and FK2 are optical phonons of the TiO$_6$ octahedron, whose ionic motions are perpendicular to the interface[24]. For FK1, the Ti atom vibrates out-of-phase with the 6 oxygen atoms; while for FK2, the Ti and 2 apical oxygen atoms vibrate in-phase with each other, but out-of-phase with the 4 in-plane oxygen atoms. As indicated by the intensities of γ' and γ*, the electrons in FeSe interact more strongly with FK1, which creates a larger electric dipole field, than with FK2.

**Superconducting gap variation**

To characterize the superconductivity, we collected extensive data on the superconducting gaps of our high-quality samples, which exhibit intriguing variations. Figure 2 shows the data measured at 6 K on six representative FeSe/ST$^{16}$O films, #1-#6, together with that of the K-dosed FeSe thick film (see Method and Supplementary for their growth and annealing conditions), all with almost identical carrier concentration between 0.11~0.12 e$^-$ per Fe evident from the similar size of the Femi surfaces (Fig. 2a). This is the typical doping for FeSe/STO, similar to those reported in films with 60~65 K $T_c$ (refs. 3,4), while the doping of the K-dosed FeSe thick film is in its optimal doping regime[14]. The Photoemission data taken across M all show Bogoliubov quasiparticle dispersions with the opening of the superconducting gaps in the two electron bands (Fig. 2b). As shown by the energy distribution curves (EDCs) at the normal state Fermi momenta ($k_F$'s) $k_1$ and $k_2$ (Fig. 2c and Supplementary Information, and the momentum locations of $k_1$ and $k_2$ are shown in Fig. 2b), the superconducting coherence peaks in these samples are sharper than those reported before; in particular, the intensity ratios between the coherence peak and the background intensity in the S16-series samples are much larger than in previous reports (gray curve in Fig.2c from ref. 3, for instance)[3-5], confirming the high quality of the films in our study. Moreover, the full widths at half maximum (FWHMs) of the coherence where $k_2$ is located. **i-k**, Same as panels (**e-g**) but for EDCs at $k_2$ and gap $\Delta_2$. All data were measured at 6 K.

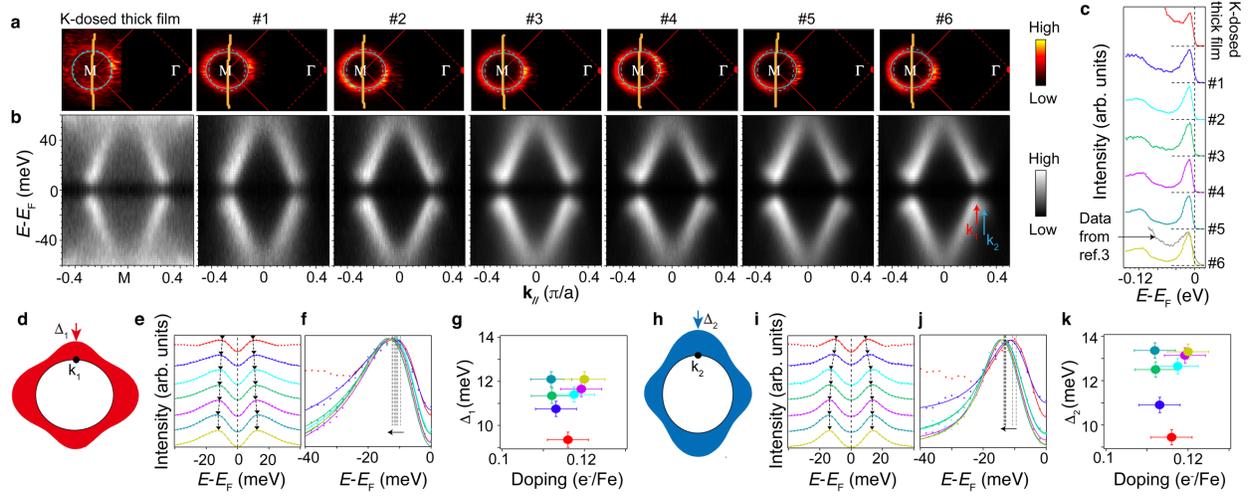

**Figure 2 Superconducting gap and doping of single layer FeSe/ST$^{16}$O. a,b**, Photoemission intensity map at the Fermi energy (**a**), and symmetrized photoemission spectra with respect to the Fermi energy (**b**) of various samples, including a K-dosed thick film and samples #1-#6. The arrows indicate the Fermi momenta $k_1$ and $k_2$. **c**, EDCs of different samples at the Fermi momentum $k_1$. All samples show greatly reduced background and sharper coherence peaks compared with the sample with T$_c$ =60±5 K in ref. 3 (grey curve). Temperature dependent studies show that the gap of S16 with $\Delta_1$=12.1meV closes at 61±3.5 K (see Supplementary Information), which is almost identical to the Tc of the film reported in ref. 3 with a much broader coherence peak. **d**, illustration of the electron pocket where $k_1$ is located and the anisotropic superconducting gap structure. **e**, Symmetrized EDCs at $k_1$ of different samples (dots) and the empirical fits (solid lines, see Supplementary Information). **f**, Zoomed-in view of fitted curves from panel (**e**) overlaid together to show the gap variation more clearly. **g**, $\Delta_1$ as a function of doping for the above samples. Doping variation is minimal, while the variation in gap sizes does not correlate with the doping. **h**, Same as panel (**d**) but of the other electron pocket,
8

peaks are almost identical for all the samples, suggesting comparable sample quality among these samples (more data are shown in the Supplementary Information). Figure 2e shows the symmetrized EDCs measured at $k_1$ for samples #1-#6 – the superconducting gap sizes $\Delta_1$ are determined by a standard empirical fit of the data to a superconducting spectral function[35-37] (see Supplementary Information). $\Delta_1$ is smallest in the K-dosed FeSe thick film, while it increases from S16 #1 to #6, as further indicated in Fig. 2f (we note that taking the coherence peak position as the superconducting gap would give the same conclusions, although the gap amplitude would be slightly larger than the fitted value). A similar trend has been observed for the superconducting gap at $k_2$ ($\Delta_2$, Figs. 2i-j). To get better statistics, the gap sizes measured at equivalent $k_F$'s are averaged, and the gap uncertainty can be reduced to 0.35 meV (see Supplementary Information for details). $\Delta_1$ varies from 9.3 meV to 12.1 meV, and $\Delta_2$ varies from 9.45 meV to 13.3 meV for different samples (Figs. 2g and 2k). Such sizable variations are significantly beyond the experimental uncertainty. Moreover, $\Delta_1$ is smaller than $\Delta_2$ in each sample. In fact, the gap anisotropy along the elliptical Fermi surfaces has been reported for FeSe/STO/KTaO$_3$ and FeSe/STO films previously, with two different local gap maxima at $k_1$ and $k_2$, and gap minima at the crossings of the two pockets[9,34], as sketched in Figs. 2d and 2h.

As shown in Figs. 2g and 2k, there is no correlation between the gap size and the doping level here. Indeed, the $T_c$ and superconducting gap are nearly constant around the optimal doping of K-dosed thick FeSe films in previous reports[14]. Therefore, we can rule out the variations in doping as the cause of the variation in gap sizes. Moreover, because our samples have almost the same coherence peak width (Fig. 2c and Supplementary Information), film quality is almost identical in these samples and its effect on the superconducting gap can be excluded here. The variation of the superconducting gaps among films with almost identical doping and sample quality thus implies that another factor must play a crucial role in the interfacial superconductivity.

**Phonon-enhanced superconductivity**

Having ruling out the most obvious causes for the variation of the superconducting gap among the six S16 films, we now examine the role interfacial EPI plays here. Recent theoretical analyses[5,25-29] show that the dimensionless electron-phonon coupling constant $\lambda$ is proportional to the ratio between the weights of the side band and the main band. Data of our S16 films clearly exhibit replica bands (Fig. 3a), and we focus on $\gamma$' here since it is more pronounced than $\gamma$*. In Fig. 3b, the integrated EDCs near M exhibit three features for the $\gamma$, $\gamma$' and $\beta$ bands, which are much more pronounced than previously reported[5,10,23]. This allows a systematic analysis and quantitative comparison on the side band intensity ratio. Following the procedure in ref. 5, we obtained the background of each EDC by cubic spline interpolation (Fig. 3b, see Supplementary Information for details). Figure 3c shows the photoemission spectra after background subtraction, in which $\gamma$' becomes more and more pronounced from sample #1 to #6, which is also clear in the EDCs before background subtraction (Fig. 3b), and after background subtraction and normalization by the peak height of the $\gamma$ band (Fig. 3d). The EPI variation in samples with subtle change in growth/annealing conditions may relate with the slightly varied interfacial bonding condition, such as the varied bond disorder between FeSe and STO (ref. 38) , which requires future studies. To quantify the EPI strength, one can define the ratio of the peak areas, $\eta=I_1/I_0$, where $I_1$ and $I_0$ are the spectral weights of $\gamma'$ and $\gamma$ after the background subtraction, respectively. Intriguingly, the interfacial EPI variation in these samples follows the variation of superconducting gap in these samples (Fig. 3e).

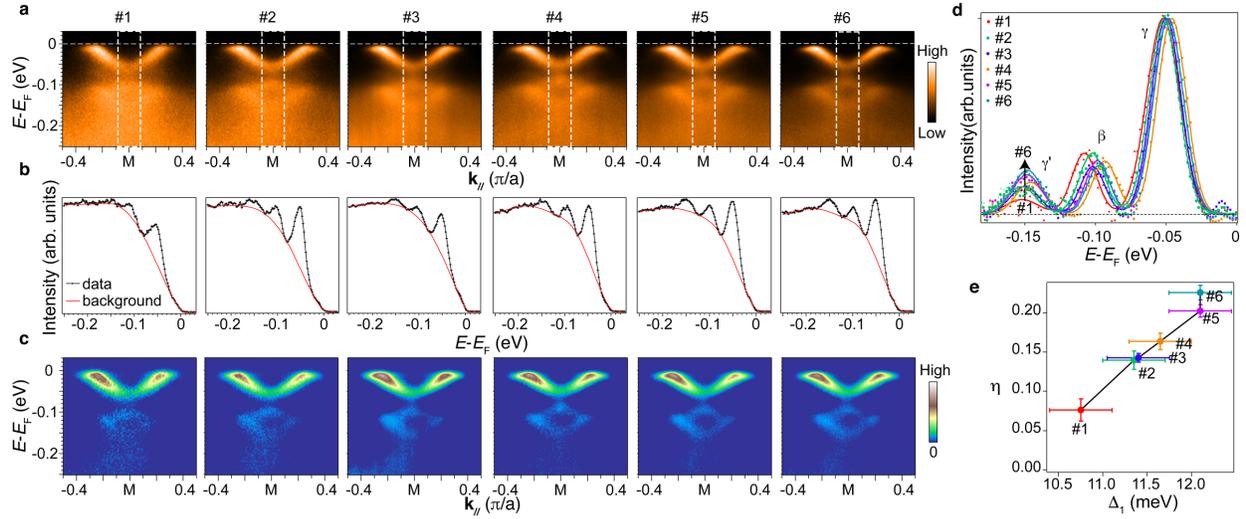

**Figure 3 Side band intensity in S16 samples. a,** Photoemission intensity distributions across M measured in S16 samples #1-#6. **b,** EDCs around M and the corresponding background used in intensity analysis for samples #1-#6. To get better statistics, the EDCs are integrated over the momentum range as indicated by the white dashed rectangle in panel a. The background is modeled using a cubic spline interpolation following the method in ref. 5 (see Supplementary Information). **c,** Background-subtracted photoemission intensity distributions of samples #1-#6 across M. **d,** Data and fits of the background-subtracted EDCs around M. The data (dots) are fitted to three Gaussian peaks, representing the spectral weight from band $\gamma$, band $\beta$, and band $\gamma'$ ($\gamma^*$ is ignored considering its low spectral weight, see Supplementary Information). The spectral weights of band $\gamma$ and band $\gamma'$ are noted as $I_0$ and $I_1$, respectively. The data and fits are normalized to the peak height of γ band. **e**, Replica band intensity ratios $\eta=I_1/I_0$ in samples #1-#6 as a function of $\Delta_1$. All data were measured at 6 K.

To further investigate the relation between superconductivity and interfacial EPI, Fig. 4 plots the superconducting gap as a function of η, which includes data points from the six representative samples, K-dosed thick FeSe film and additional high-quality samples (see Supplementary Information for spectra of these samples). Our data demonstrate a linear dependence of $\Delta_1$ and $\Delta_2$ on the relative intensity of the side band. BCS theory predicts an exponential relation between superconducting gap and the electron-phonon coupling. However, an exponential fit to our data gives an unrealistically small phonon frequency (see Supplementary Information). On the other hand, the linear increase of gap with $\lambda$ is a hallmark of the pairing-enhancement scenario where electron-phonon interactions are strongly peaked in the forward scattering (q=0) direction[5,25-29].

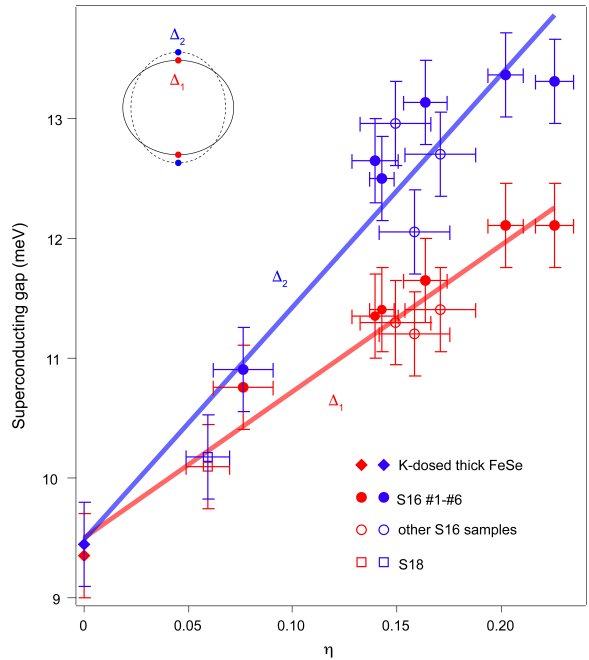

**Figure 4 Superconducting gap as a function of electron-phonon coupling strength.** Superconducting gap sizes $\Delta_1$ (red) and $\Delta_2$ (blue) are plotted as a function of the intensity ratio

between the side band and the main band, η, which indicates the electron-phonon coupling strength. The red and blue bars are the linear fits to $\Delta_1$ and $\Delta_2$, respectively. Note: here we use $\eta=I_1/I_0$ to represents the electron phonon coupling constant according to theory[5,29], while using $I_1/(I_1+I_0)$ would only slightly change the *x* axis and does not affect the conclusion.

The extrapolation to the η = 0 limit of the linear relations of both $\Delta_1$ and $\Delta_2$ give an intercept around 9.5meV (Fig. 4), which coincides with the superconducting gap of a heavily-electron-doped FeSe monolayer on the FeSe thick film[14], where the heavy electron doping comes from the K atoms dosed on the surface. That is, in the absence of interfacial electron-phonon coupling, FeSe/STO is similar to a K-dosed surface FeSe layer of a thick film, whose intrinsic pairing mechanism, likely spin/orbital fluctuations, can generate a sizeable superconducting gap already, while the interfacial EPI accounts for a linear enhancement on top of that. Our data thus directly prove that the high $T_c$ in FeSe/STO is caused by the collaboration between an intrinsic mechanism and the interfacial electron-phonon interactions[5,25,26]. This is consistent with the contention that although such forward-scattering EPI can effectively enhance superconductivity, the EPI strength is not sufficient to produce the high $T_c$ in FeSe/STO (ref. 27). The EPI is only responsible for a gap enhancement of about 3~4 meV for the FeSe/STO film with the highest $T_c$, and thus the oxygen isotope effect on the gap (presumably ~6%) would be around 0.2meV, which is beyond the current experimental uncertainty.

**Discussions and conclusions**

Our findings also set constraints for theories on forward scattering enhanced Cooper pairing. EPI would renormalize the energy of the main band and side band, and thus their separation, $E_S$, would be larger than the bare phonon energy. This separation is predicted to be $(1+2\lambda+O(\lambda^2))\Omega_{ph}$ for forward scattering, where $\Omega_{ph}$ is the optical phonon energy[29]. Analysis of the EELS phonon linewidth gives $\lambda\sim0.25$ for the FK1 phonon in FeSe/STO[24], and $\lambda$ in the range of 0.16~0.25 is used in various theoretical calculations to produce a sizable enhancement in $T_c$ (refs. 5, 29). Consequently, for $\Omega_1 \sim94$ meV, the current theories would predict $E_S$ to be a few tens of meV higher than $\Omega_1$ (ref. 5). However, our combined study with *in-situ* ARPES and *ex-situ* EELS suggest that the average $E_S$ is only a few meV larger than $\Omega_1$, while the difference between $E_S^*$ and $\Omega_2$ is almost negligible, as its coupling is weaker. Consistently, for our S16 #1-#6 films, although η (thus $\lambda$) doubles, $E_S$ only varies by less than 3 meV. Although strictly speaking, a quantitative study requires *in-situ* ARPES and *in-situ* EELS on the same sample, our data demonstrate that the $\lambda$ necessary for the $T_c$ enchantment and/or the renormalization of the electronic bands by the forward scattering EPI might be greatly overestimated in the current theoretical framework, which calls for further study.

In a recent scanning tunneling spectroscopy study, the pairing symmetry of FeSe/STO was identified to be plain *s*-wave[39], which would be compatible with a further enhancement by interfacial electron-phonon interactions. On the other hand, it has been shown that electron-phonon forward scattering can also collaborate with various spin and orbital fluctuations to enhance superconductivity, be it *s*-wave or *d*-wave pairing[25] etc. This suggests that electron-phonon forward scattering mechanism can be applied to a broad range of superconducting materials[40]. For example, there are oxide layers in many cuprate and iron-based superconductors, it would be intriguing to search for such effects.

In summary, our superior quality films and high quality ARPES data enable the quantitative examination of the relationship between the superconductivity and interfacial electron phonon interactions in FeSe/STO. Our experiments constitute direct evidence that the greatly enhanced superconductivity in FeSe/STO is the combined effects of two mechanisms, one intrinsic to heavily electron-doped FeSe, and the other an additional enhancement of superconductivity that depends linearly on the strength of the interfacial electron-phonon coupling. Our findings compellingly solve the mystery of why $T_c$ is so high in FeSe/STO, and set constraints on current theories. Moreover, our results directly establish that electron-phonon interactions, particularly the forward-scattering type, can play a critical role in the high $T_c$ of a highly correlated superconductor, suggesting a route forward for the development of interfacial high-$T_c$ superconductors and the understanding of high-$T_c$ superconductivity in general.

## Methods

**Oxygen isotope substitution and sample preparation**

The method to grow $SrTi^{18}O_3$ films atop commercial $SrTi^{16}O_3$ substrates, and the secondary ion mass spectrometry (SIMS) results that confirm the isotope substitution, are described in the Supplementary Information. To obtain $SrTi^{16}O_3/SrTi^{16}O_3$ with identical surface quality and oxygen vacancy concentration as $SrTi^{18}O_3/SrTi^{16}O_3$, we anneal the substrates, grow 60 unit cells of $SrTi^{16}O_3$, then anneal these under the same conditions as $SrTi^{18}O_3/SrTi^{16}O_3$ but in an $^{16}O_2$ atmosphere. After the preparation of the $SrTi^{18}O_3$ or $SrTi^{16}O_3$ surface, the samples are transferred under ultra-high vacuum for FeSe growth. Single-layer FeSe films were grown at ~520 ℃ by co-evaporation of Se and Fe and then post-annealed at ~546 ℃ for 5.5-8.5 hours (see Supplementary Information for more details). Thick FeSe films were grown at 370 ℃ then post-annealed at 410 ℃ in vacuum for 2.5 hours. Surface K dosing is conducted with a commercial SAES alkali dispenser.

**ARPES measurements**

The in-house ARPES measurements were performed with a Fermi Instruments discharge lamp (21.2 eV He-Iα light) and a Scienta DA30 electron analyzer. The overall energy resolution is 7.5 meV, and the angular resolution is 0.3°. Samples were measured under an ultra-high-vacuum of $5\times10^{-11}$ torr. The sample growth, K dosing and ARPES measurements were all conducted *in situ*.

**EELS measurements**

Single-layer FeSe/SrTi$^{18}$O$_3$/SrTi$^{16}$O$_3$ and single-layer FeSe/SrTi$^{16}$O$_3$/SrTi$^{16}$O$_3$ samples were capped by amorphous Se to protect the surface from atmosphere. The capped samples were transferred to a high-resolution EELS system, and annealed at 450 ℃ for 6 hours to remove the Se capping layer. LEED patterns were collected to confirm the removal of the capping layer and verify the sample quality. High-resolution EELS measurements were performed at 35K, with an incident beam energy of 110 eV and an incident angle of 60° with respect to the surface normal. The energy resolution is 3 meV.

**Supplementary Information** is available in the online version of the paper.

**Acknowledgements** This work is supported in part by the National Science Foundation of China, National Key R&D Program of the MOST of China (Grant Nos. 2016YFA0300200, 2017YFA0303004), the Science and Technology Committee of Shanghai under the grant Nos.

15YF1401000 and 15ZR1402900. We appreciate the helpful discussions with Profs. Dragan Mihailovic, Hugo Keller, and Annette Bussmann-Holder, Frederick Walker, Victor E. Henrich on the isotope effects, and helpful discussions with Profs. Guangming Zhang and Qianghua Wang on EPI. We thank Dr. Darren Peets for helping with the editing, and Dr. Qiuyun Chen for helping with the SIMS measurements.

**Author contributions** Q.S. T.L.Y., X.L., C.H.P.W., Q.Y, B.P.X. and H.C.X. conducted film growth and ARPES measurements. R.P., Q.S. T.L.Y. and D.L.F. analyzed the ARPES data. S.Y.Z, X.T.Z and J.D.G. conducted EELS measurements. R.P. and D.L.F wrote the paper. D.L.F. is responsible for project direction, planning and infrastructure.

**Author Information** The authors declare no competing financial interests. Correspondence and requests for materials should be addressed to D.L.F. (dlfeng@fudan.edu.cn), R.P. (pengrui@fudan.edu.cn).